\documentclass[12pt, a4paper]{article}
\usepackage{graphicx}
\usepackage{latexsym}
\usepackage{amsmath}
\usepackage{amssymb}
\usepackage{epsfig}
\begin{document}

\title{Dark Energy:  Recent Developments\thanks{Invited ``brief review'' for
Modern Physics Letters A; to appear.}}

\author{Norbert Straumann\\
Institute for Theoretical Physics University of Zurich,\\
Winterthurerstrasse 180, CH--8057 Zurich, Switzerland}
\maketitle

\begin{abstract}
A six parameter cosmological model, involving a vacuum energy
density that is extremely tiny compared to fundamental particle
physics scales, describes a large body of increasingly accurate
astronomical data. In a first part of this brief review we summarize
the current situation, emphasizing recent progress. An almost
infinitesimal vacuum energy is only the simplest candidate for a
cosmologically significant nearly homogeneous exotic energy density
with negative pressure, generically called Dark Energy. If general
relativity is assumed to be also valid on cosmological scales, the
existence of such a dark energy component that dominates the recent
universe is now almost inevitable. We shall discuss in a second part
the alternative possibility that general relativity has to be
modified on distances comparable to the Hubble scale. It will turn
out that observational data are restricting theoretical speculations
more and more. Moreover, some of the recent proposals have serious
defects on a fundamental level (ghosts, acausalities, superluminal
fluctuations).
\end{abstract}

\newpage

\section{Introduction}

On the basis of a rich body of astronomical observations there is
now convincing evidence that the recent ( $z<1$) Universe is
dominated by an exotic nearly homogeneous dark energy density with
{\it negative} pressure. The simplest candidate for this unknown
so-called \emph{Dark Energy} (DE) is a cosmological term in
Einstein's field equations, a possibility that has been considered
during all the history of relativistic cosmology. Independently of
what this exotic energy density is, one thing is certain since a
long time: The energy density belonging to the cosmological constant
is not larger than the cosmological critical density, and thus
\emph{incredibly small by particle physics standards}. This is a
profound mystery, since we expect that all sorts of \emph{vacuum
energies} contribute to the effective cosmological constant.

Since this is such an important issue for fundamental physics,
astrophysics and cosmology, it should be of interest to indicate how
convincing the evidence for this finding really is, or whether one
should still remain sceptical. Much of this is based on the observed
temperature fluctuations of the cosmic microwave background
radiation (CMB), and large scale structure formation. When combined
with other measurements a cosmological world model of the
Friedmann-Lema\^{\i}tre variety has emerged that is spatially almost
flat, with about 70\% of its energy contained in the form of Dark
Energy.

\section{Luminosity-Redshift Relation of Type Ia Supernovae}

The first serious evidence for a currently accelerating universe,
and still the only direct one, came from the Hubble diagram for Type
Ia supernovae, that are good -- although not perfect -- standard
candles.

In an ideal Friedmann-Lema\^{\i}tre universe, it is easy to
establish a relationship between the luminosity distance, $D_L$, of
an ideal standard candle and and the redshift, $z$, of the source.
We recall that $D_L$ is defined by $D_L =
(\mathcal{L}/4\pi\mathcal{F})^{1/2}$, where $\mathcal{L}$ is the
intrinsic luminosity of the source and $\mathcal{F}$ the observed
energy flux. Astronomers use as logarithmic measures of
$\mathcal{L}$ and $\mathcal{F}$ the {\it absolute and apparent
magnitudes}\footnote{Beside the (bolometric) magnitudes $m,M$,
astronomers also use magnitudes $m_B,~m_V,~\ldots$ referring to
certain wavelength bands $B$ (blue), $V$ (visual), and so on.},
denoted by $M$ and $m$, respectively. The conventions are chosen
such that the {\it distance modulus} $m-M$ is related to $D_L$ as
follows
\begin{equation}
m-M = 5 \log \left( \frac{D_L}{1~Mpc}\right) + 25. \label{eq:SN24}
\end{equation}

With the help of the Freedmann equations one can express the product
of the Hubble parameter, $H_0$, and $D_L$ as a function of $z$ and
the cosmological density parameters $\Omega_X$ for the various
species, $X$, of the energy-matter content, including  Dark Energy.
The comparison of the resulting theoretical {\it magnitude -
redshift relation} with data leads to interesting restrictions for
the cosmological $\Omega$-parameters. In practice often only
$\Omega_M$ and $\Omega_\Lambda$, the density corresponding to the
cosmological constant $\Lambda$, are kept as independent parameters,
where from now on the subscript $M$ denotes non-relativistic (mostly
cold dark) matter.

In view of the complex physics involved, it is not astonishing that
type Ia  supernovas are not perfect standard candles. Their peak
absolute magnitudes have a dispersion of 0.3 - 0.5 mag, depending on
the sample. Astronomers have, however, learned in recent years to
reduce this dispersion by making use of empirical correlations
between the absolute peak luminosity and light curve shapes.
Examination of nearby SNe showed that the peak brightness is
correlated with the time scale of their brightening and fading: slow
decliners tend to be brighter than rapid ones. Using these and other
correlations it became possible to reduce the remaining intrinsic
dispersion, at least in the average, to $\simeq 0.15 mag$. (For the
various methods in use, and how they compare, see e.g. \cite{leib1},
\cite{rie}, and references therein.) Other corrections, such as
Galactic extinction, have been applied, resulting for each supernova
in a corrected (rest-frame) magnitude.

\begin{figure}
\begin{center}
\includegraphics[height=0.3\textheight]{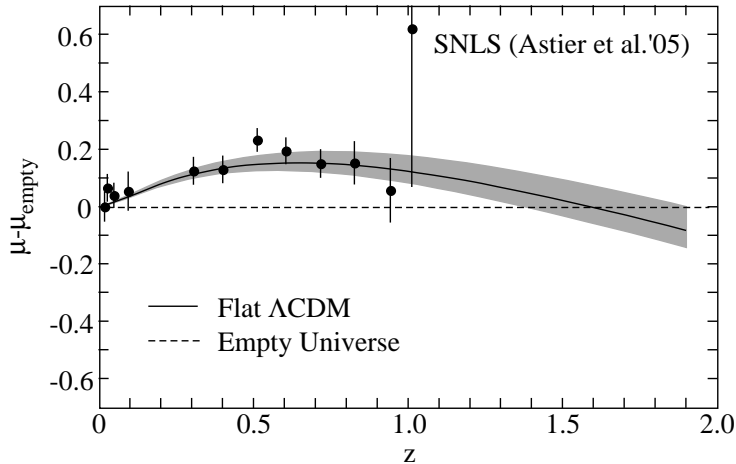}
\caption{Distance moduli relative to an empty uniformly expanding
universe (residual Hubble diagram) for SNe Ia of the SNLS data
\cite{leg}. The shaded area shows the prediction for the
luminosity-redshift relation from the $\Lambda$CDM model model fit
to the three-year WMAP data only. (From Fig.~8 of
Ref.~\cite{Sperg2}.) \label{cosmp:Fig-3}}
\end{center}
\end{figure}

After the classic papers \cite{p99},~\cite{s98},~\cite{r98} on the
Hubble diagram for high-redshift Type Ia supernovae, published by
the SCP and HZT teams, significant progress has been made (for
reviews, see Refs.~\cite{leib2}, \cite{fil04}). The results,
presented in Ref.~\cite{rie}, are based on additional data for
$z>1$, obtained in conjunction with the GOODS (Great Observatories
Origins Deep Survey) Treasury program, conducted with the Advanced
Camera for Surveys (ACS) aboard the Hubble Space Telescope (HST). In
the meantime new results have been published. Perhaps the best
high-$z$ SN Ia compilation to date are the results from the
Supernova Legacy Survey (SNLS) of the first year \cite{leg}. The
other main research group has also published new data at about the
same time \cite{clo}. Fig.~\ref{cosmp:Fig-3} shows the data points
of Ref.~\cite{leg} for the distance moduli relative to an empty
uniformly expanding universe as a function of redshift. Also shown
is the prediction of the best fit values of a six parameter
$\Lambda$CDM model, using only the three-year WMAP data (see
Sect.~4).

Possible systematic uncertainties due to astrophysical effects have
been discussed extensively in the literature. The most serious ones
are (i) {\it dimming} by intergalactic dust, and (ii) {\it
evolution} of SNe Ia over cosmic time, due to changes in progenitor
mass, metallicity, and C/O ratio.

To improve the observational situation a satellite mission called
SNAP (``Supernovas Acceleration Probe'') has been proposed
\cite{snap}. According to the plans this satellite would observe
about 2000 SNe within a year and much more detailed studies could
then be performed. For the time being some scepticism with regard to
the results that have been obtained is still not out of place, but
the situation is steadily improving.

Finally, we point out a more theoretical complication. In the
analysis of the data the luminosity distance for an ideal Friedmann
universe was always used. But the data were taken in the real
inhomogeneous Universe. The magnitude-redshift relation in a
perturbed Friedmann model has been derived in \cite{Sas}, and was
later used to determine the angular power spectrum of the luminosity
distance (the $C_l$'s defined in analogy to (\ref{eq:CMB1}))
\cite{SSS}. One of the numerical results was that the uncertainties
in determining cosmological parameters via the magnitude-redshift
relation caused by fluctuations are small compared with the
intrinsic dispersion in the absolute magnitude of Type Ia
supernovae.

\section{Microwave Background Anisotropies}

Investigations of the cosmic microwave background have presumably
contributed most to the remarkable progress in cosmology during
recent years. Beside its spectrum, which is Planckian to an
incredible degree, we also can study the temperature fluctuations
over the ``cosmic photosphere'' at a redshift $z\approx1100$.
Through these we get access to crucial cosmological information
(primordial density spectrum, cosmological parameters, etc). A major
reason for why this is possible relies on the fortunate circumstance
that the fluctuations are tiny ($\sim 10^{-5}$ ) at the time of
recombination. This allows us to treat the deviations from
homogeneity and isotropy for an extended period of time
perturbatively, i.e., by linearizing the Einstein and matter
equations about solutions of the idealized Friedmann-Lema\^{\i}tre
models. Since the physics is effectively {\it linear}, we can
accurately work out the {\it evolution} of the perturbations during
the early phases of the Universe, given a set of cosmological
parameters. Confronting this with observations, tells us a lot about
the cosmological parameters as well as the initial conditions, and
thus about the physics of the very early Universe. Through this
window to the earliest phases of cosmic evolution we can, for
instance, test general ideas and specific models of inflation.

\subsection{Qualitative remarks}

We begin with some qualitative remarks. Long before recombination
(at temperatures $T>6000 K$, say) photons, electrons and baryons
were so strongly coupled that these components may be treated
together as a single fluid. In addition to this there is also a dark
matter component. For all practical purposes the two interact only
gravitationally. The investigation of such a two-component fluid for
small deviations from an idealized Friedmann behavior is a
well-studied application of cosmological perturbation theory (see,
e.g., Ref.~\cite{NS8}).

At a later stage, when decoupling is approached, this approximate
treatment breaks down because the mean free path of the photons
becomes longer (and finally `infinite' after recombination). While
the electrons and baryons can still be treated as a single fluid,
the photons and their coupling to the electrons have to be described
by the general relativistic Boltzmann equation. The latter is, of
course, again linearized about the idealized Friedmann solution.
Together with the linearized fluid equations (for baryons and cold
dark matter, say), and the linearized Einstein equations one arrives
at a complete system of equations for the various perturbation
amplitudes of the metric and matter variables. There exist widely
used codes, e.g. CMBFAST \cite{22}, that provide the CMB
anisotropies -- for given initial conditions -- to a precision of
about 1\%. A lot of qualitative and semi-quantitative insight into
the relevant physics can, however, be gained by looking at various
approximations of the basic dynamical system.

Let us first discuss the temperature fluctuations. What is observed
is the temperature autocorrelation:
\begin{equation}
 C(\vartheta ):= \left\langle \frac{\Delta T(\mathbf{n})}{T}\cdot
\frac{\Delta T(\mathbf{n'})}{T}\right\rangle\\
= \sum_{l=2}^\infty \frac{2l+1}{4\pi} C_l P_l(\cos \vartheta) ,
\label{eq:CMB1}
\end{equation}
where $\vartheta$ is the angle between the two directions of
observation $\mathbf{n}, \mathbf{n'}$, and the average is taken
ideally over all sky. The {\it angular power spectrum} is by
definition $[l(l+1)/2\pi]C_l \; \; versus \; \;l \; \; (\vartheta
\simeq \pi /l ).$

A characteristic scale, which is reflected in the observed CMB
anisotropies, is the sound horizon at last scattering, i.e., the
distance over which a pressure wave can propagate until decoupling.
This can be computed within the unperturbed model and subtends about
half a  degree on the sky for typical cosmological parameters. For
scales larger than this sound horizon the fluctuations have been
laid down in the very early Universe. These have been detected by
the COBE satellite. The (gauge invariant brightness) temperature
perturbation $\Theta = \Delta T/T$ is dominated by the combination
of the intrinsic temperature fluctuations and gravitational redshift
or blueshift effects. For example, photons that have to climb out of
potential wells for high-density regions are redshifted. One can
show that these effects combine for adiabatic initial conditions to
$\frac{1}{3}\Psi$, where $\Psi$ is one of the two gravitational
Bardeen potentials. The latter, in turn, is directly related to the
density perturbations. For scale-free initial perturbations and
almost vanishing spatial curvature the corresponding angular power
spectrum of the temperature fluctuations turns out to be nearly flat
(Sachs-Wolfe plateau).

On the other hand, inside the sound horizon before decoupling,
acoustic, Doppler, gravitational redshift, and photon diffusion
effects combine to the spectrum of small angle anisotropies shown in
Figure \ref{cosmp:Fig-1}. These result from gravitationally driven
synchronized acoustic oscillations of the photon-baryon fluid, which
are damped by photon diffusion.

A particular realization of $\Theta(\mathbf{n})$, such as the one
accessible to us (all sky map from our location), cannot be
predicted. Theoretically, $\Theta $ is a random field,
$\Theta(\mathbf{x},\eta,\mathbf{n})$, depending on the conformal
time $\eta$, the spatial coordinates $\mathbf{x}$, and the observing
direction $\mathbf{n}$. Its correlation functions should be
rotationally invariant in $\mathbf{n}$, and respect the symmetries
of the background time slices. If we expand $\Theta$ in terms of
spherical harmonics,
\begin{equation}
\Theta(\mathbf{n}) = \sum_{lm} a_{lm} Y_{lm}(\mathbf{n}),
\label{eq:CMB2}
\end{equation}
the random variables $a_{lm}$ have to satisfy
\begin{equation}
\langle a_{lm}\rangle = 0,\;\;\;  \langle a_{lm}^\star
a_{l'm'}\rangle = \delta_{ll'}\delta_{mm'}C_l(\eta), \label{eq:CMB3}
\end{equation}
where the $C_l(\eta)$ depend only on $\eta$. Hence the correlation
function at the present time $\eta_0$ is given by (\ref{eq:CMB1}),
where $C_l = C_l(\eta_0)$, and the bracket now denotes the
statistical average. Thus,
\begin{equation}
C_l = \frac{1}{2l+1}\left\langle\sum_{m=-l}^la_{lm}^\star
a_{lm}\right\rangle. \label{eq:CMB4}
\end{equation}
The standard deviations $\sigma(C_l)$ measure a fundamental
uncertainty in the knowledge we can get about the $C_l$'s. These are
called {\it cosmic variances}, and are most pronounced for low $l$.
In simple inflationary models the $a_{lm}$ are Gaussian distributed,
hence
\begin{equation}
\frac{\sigma(C_l)}{C_l} = \sqrt{\frac{2}{2l+1}}. \label{eq:CMB5}
\end{equation}
Therefore, the limitation imposed on us (only one sky in one
universe) is small for large $l$.

A polarization map of the CMB radiation provides important
additional information to that obtainable from the temperature
anisotropies. For example, we can get constraints about the epoch of
reionization. Most importantly, future polarization observations may
reveal a stochastic background of gravity waves, generated in the
very early Universe. The polarization tensor of an all sky map of
the CMB radiation can be parametrized in temperature fluctuation
units, relative to the orthonormal basis $\{d\vartheta,
\sin\vartheta\; d\varphi\}$ of the two sphere, in terms of the Pauli
matrices as $\Theta\cdot 1 + Q\sigma_3 + U\sigma_1 + V\sigma_2$. The
Stokes parameter $V$ vanishes (no circular polarization). Therefore,
the polarization properties can be described by a symmetric
trace-free tensor on $S^2$. As for gravity waves, the components $Q$
and $U$ transform under a rotation of the 2-bein by an angle
$\alpha$ as $Q \pm iU \rightarrow e^{\pm 2i\alpha}(Q\pm iU)$, and
are thus of spin-weight 2. ``Electric'' and ``magnetic'' multipole
moments are defined by the decomposition
\begin{equation}
Q+iU = \sqrt{2}\sum_{l=2}^\infty\sum_{m}\left[a^E_{(lm)}+
ia^B_{(lm)}\right] \,_{2\!}Y_l^m, \label{eq:CMB14}
\end{equation}
where $\,_{s\!}Y_l^m$ are the spin-s harmonics.

As in Eq.~(\ref{eq:CMB2}) the multipole moments $a^E_{(lm)}$ and
$a^B_{(lm)}$ are random variables and determine, similar to
(\ref{eq:CMB1}) and (\ref{eq:CMB4}), the various angular correlation
functions.

\section{Observational Results and Cosmological\\ Parameters}

In recent years several experiments gave clear evidence for multiple
peaks in the angular temperature power spectrum at positions
expected on the basis of the simplest inflationary models and big
bang nucleosynthesis\cite{25}. These results have been confirmed and
substantially improved by the first year WMAP data \cite{26},
\cite{27}. Fortunately, the improved data after three years of
integration are now available \cite{Sperg2}. Below we give a brief
summary of some of the most important results.

\begin{figure}
\begin{center}
\includegraphics[height=0.3\textheight]{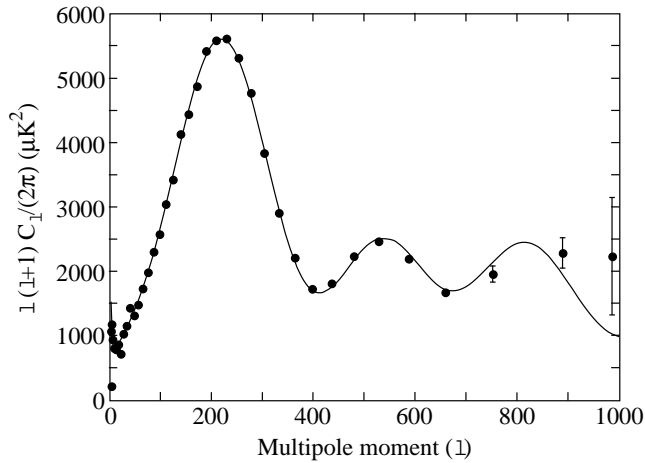}
\caption{Three-year WMAP data for the temperature-temperature (TT)
power spectrum. The black line is the best fit $\Lambda$CDM model
for the three-year WMAP data. (Adapted from Figure 2 of
Ref.~\cite{Sperg2}.) \label{cosmp:Fig-1}}
\end{center}
\end{figure}

\begin{figure}
\begin{center}
\includegraphics[height=0.3\textheight]{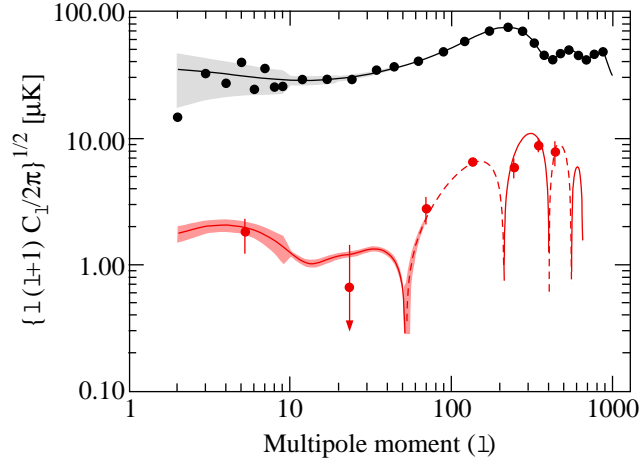}
\caption{WMAP data for the temperature-polarization TE power
spectrum. The best fit $\Lambda$CDM model is also shown. (Adapted
from Figure 25 of Ref.~\cite{Pag}.) \label{cosmp:Fig-2}}
\end{center}
\end{figure}

Fig.~\ref{cosmp:Fig-1} shows the 3-year data of WMAP for the TT
angular power spectrum, and the best fit (power law) $\Lambda$CDM
model. The latter is a spatially flat model and involves the
following six parameters: $\Omega_bh^2,~\Omega_Mh^2,~H_0$, amplitude
of fluctuations, $\sigma_8$, optical depth, $\tau$, and the spectral
index, $n_s$, of the primordial scalar power spectrum.
Fig.~\ref{cosmp:Fig-2} shows in addition the TE polarization data
\cite{Pag}. There are now also EE data that lead to a further
reduction of the allowed parameter space. The first column in Table
1 shows the best fit values of the six parameters, using only the
WMAP data.

\begin{figure}
\begin{center}
\includegraphics[height=0.3\textheight]{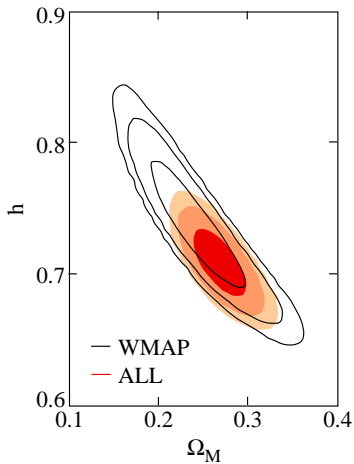}
\caption{Joint marginalized contours (68\% and 95\% confidence
levels) in the ($\Omega_M,h$)-plane for WMAP only (solid lines) and
additional data (filled red) for the power-law $\Lambda$CDM model.
(From Fig.~10 in \cite{Sperg2}.) \label{cosmp:Fig-4}}
\end{center}
\end{figure}

Combining the WMAP results with other astronomical data reduces the
uncertainties for some of the six parameters. This is illustrated in
the second column which shows the 68\% confidence ranges of a joint
likelihood analysis when the power spectrum from the completed
2dFGRS \cite{Col} is added. In Ref.~\cite{Sperg2} other joint
constraints are listed (see their Tables 5, 6). In
Fig.~\ref{cosmp:Fig-4} we reproduce one of many plots in
\cite{Sperg2} that shows the joint marginalized contours in the
($\Omega_M,h$)-plane.

The parameter space of the cosmological model can be extended in
various ways. Because of intrinsic degeneracies, the CMB data alone
no more determine unambiguously the cosmological model parameters.
We illustrate this for non-flat models. For these the WMAP data (in
particular the position of the first acoustic peak) restricts the
curvature parameter $\Omega_K$ to a narrow region around the
degeneracy line $\Omega_K=-0.3040+0.4067~\Omega_\Lambda$. This does
not exclude models with $\Omega_\Lambda=0$. However, when for
instance the Hubble constant is restricted to an acceptable range,
the universe must be nearly flat. For example, the restriction
$h=0.72\pm 0.08$ implies that $\Omega_K=-0.003^{+0.013}_{-0.017}$
and $\Omega_\Lambda= 0.758^{+0.035}_{-0.058}$. Other strong limits
are given in Table 11 of Ref.~\cite{Sperg2}, assuming that the
equation of state parameter, $w$, has the value $-1$ of vacuum
energy. But even when this is relaxed, the combined data constrain
$\Omega_K$ and $w$ significantly (see Figure 17 of \cite{Sperg2}).
The marginalized best fit values are
$w=-1.062^{+0.128}_{-0.079},~\Omega_K=-0.024^{+0.016}_{-0.013}$ at
the 68\% confidence level.

\vspace{1cm}

\begin{tabular}{|l||c|r|}
\multicolumn{3}{c}{\textbf{Table 1.}} \\  \hline Parameter & WMAP
alone & WMAP + 2dFGRS\\ \hline\hline 
$100\Omega_b h^2$ & $2.233^{+0.072}_{-0.091}$ &
$2.223^{+0.066}_{-0.083}$\\
$\Omega_M h^2$ & $0.1268^{+0.0072}_{-0.0095}$   & $0.1262^{+0.0045}_{-0.0062}$   \\
$h $ &   $0.734^{+0.028}_{-0.038}$   &  $0.732^{+0.018}_{-0.025}$  \\
$\Omega_M$  &  $0.238^{+0.030}_{-0.041}$   &  $0.236^{+0.016}_{-0.029}$  \\
$\sigma_8$  &  $0.744^{+0.050}_{-0.060}$   &  $0.737^{+0.033}_{-0.045}$ \\
$\tau$      &  $0.088^{+0.028}_{-0.034}$   &  $0.083^{+0.027}_{-0.031}$ \\
$n_s$       &  $0.951^{+0.015}_{-0.019}$   &  $0.948^{+0.014}_{-0.018}$ \\
\hline
\end{tabular}

\vspace{1cm}

\begin{figure}[tb]
\begin{center}
\includegraphics[height=0.3\textheight]{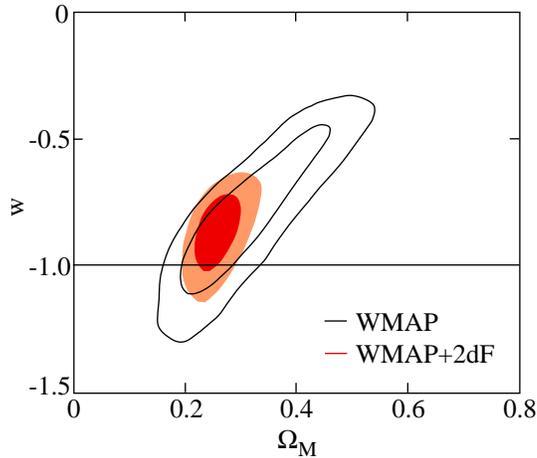}
\caption{Constraints on the equation of state parameter $w$ in a
flat universe model when WMAP data are combined with the 2dFGRS
data. (From Fig.~15 in \cite{Sperg2}.) \label{cosmp:Fig-5}}
\end{center}
\end{figure}

The restrictions on $w$ -- assumed to have no $z$-dependence -- for
a flat model are illustrated in Fig.~\ref{cosmp:Fig-5} .

Another interesting result is that reionization of the Universe has
set in at a redshift of $z_r = 10.9^{+2.7}_{-2.3}$. Later
(Sect.~6.1) we shall add some remarks on what has been learnt about
the primordial power spectrum.

It is most remarkable that a six parameter cosmological model is
able to fit such a rich body of astronomical observations. There
seems to be little room for significant modifications of the
successful $\Lambda$CDM model. An exciting result is that the WMAP
data match the basic inflationary predictions, and are even well fit
by the simplest model $V\propto \varphi^2$ (see Sect. 6 of
\cite{Sperg2}).

\section{Dynamical Models of Dark Energy}

If the vacuum energy constitutes the missing two thirds of the
average energy density of the \emph{present} Universe, we would be
confronted with the following \emph{cosmic coincidence} problem:
Since the vacuum energy density is constant in time -- at least
after the QCD phase transition --, while the matter energy density
decreases as the Universe expands, it would be more than surprising
if the two are comparable just at about the present time, while
their ratio was tiny in the early Universe and would become very
large in the distant future.  The goal of dynamical models of Dark
Energy is to avoid such an extreme fine-tuning. The ratio
$w:=p/\rho$ of this component then becomes a function of redshift.

In a large class of dynamic dark energy models the exotic missing
energy with negative pressure is described by a scalar field, whose
potential is chosen such that the energy density  of the homogeneous
scalar field adjusts itself to be comparable to the matter density
today for quite generic initial conditions, and is dominated by the
potential energy. This ensures that the pressure becomes
sufficiently negative. It is not simple to implement this general
idea such that the model is phenomenologically viable.

For an extensive recent review that contains a description of a
variety of scalar field models, see Ref.~\cite{CST}. It has to be
emphasized that on the basis of the vacuum energy problem we would
expect a huge additive constant for the quintessence potential that
would destroy the hole picture. Thus, assuming for instance that the
potential approaches zero as the scalar field goes to infinity, has
(so far) no basis. Apart of this and other fine tuning problems, I
doubt that this kind of phenomenological models -- with no natural
field theoretical justification -- will lead to an understanding of
Dark Energy at a deeper level.

\section{Alternatives to Dark Energy}
In the previous sections we have discussed some of the wide range of
astronomical data that support the following `concordance model':
The Universe is spatially flat and dominated by a Dark Energy
component and weakly interacting cold dark matter. Furthermore, the
primordial fluctuations are adiabatic, nearly scale invariant and
Gaussian, as predicted in simple inflationary models. It is very
likely that the present concordance model will survive
phenomenologically.

A dominant Dark Energy component with density parameter $\simeq 0.7$
is so surprising that it should be examined whether this conclusion
is really unavoidable. In what follows I shall briefly discuss some
alternatives that have been proposed.

\subsection{Changes in the initial conditions}

Since we do not have a tested theory predicting the spectrum of
primordial fluctuations, it appears reasonable to consider a wider
range of possibilities than simple power laws. An instructive
attempt in this direction was made some time ago \cite{29}, by
constructing an Einstein-de Sitter model with $\Omega_\Lambda = 0$,
fitting the CMB data as well as the power spectrum of 2dFGRS. In the
meantime, significant improvements in astronomical data sets have
been made. In particular, the analysis of the three year WMAP data
showed that there are no significant features in the primordial
curvature fluctuation spectrum (see Sect.~5 of Ref.~\cite{Sperg2}).
With the larger samples of high redshift supernovae and more precise
information on large scale galaxy clustering, such models with
vanishing Dark Energy are no more possible \cite{BDRS}.

\subsection{Inhomogeneous models}

\subsubsection{Back reaction}

It has recently been suggested \cite{Kol}, \cite{Kol2} that large
scale perturbations  may cause a large backreaction that could mimic
dark energy and induce acceleration. The authors stressed that for
investigating the effective dynamics averaging over a volume of size
comparable with the present-day Hubble volume is essential. To
decide on the basis of detailed calculations whether this is indeed
possible is a very difficult task. However, from what we know about
the CMB radiation it appears unlikely that there are such sizable
perturbations out to very large scales.

The work by Kolb et al.~\cite{Kol},~\cite{Kol2} triggered a lot of
activity. We add some remarks about the ongoing discussion.

\subsubsection{Exact inhomogeneous model studies}

Effects of inhomogeneous matter distribution on light propagation
were recently studied in the Lema\^{\i}tre-Tolman (LT) model, in
order to see whether these can mimic an accelerated expansion.

The LT model is a family of spherically symmetric dust solutions of
Einstein's equations. For these the magnitude-redshift relation can
be worked out exactly.

As an example we mention Ref.~\cite{Bol}, where it was shown that
for $\Lambda=0$ the observed behavior of supernovae brightness can
not be fitted, unless our position in the model universe is very
special. In that case one has to analyze also other data, in
particular the CMB angular power spectrum. At the time of writing,
this has not yet been done, but is certainly underway.

\subsection{Modifications of gravity}

Since no satisfactory explanation of Dark Energy has emerged so far,
possible modifications of GR, that would change the late expansion
rate of the universe, have recently come into the focus of
attention. The cosmic speed-up might, for instance, be explained by
sub-dominant terms (like $1/R$) that become essential at small
curvature. Modified gravity models have to be devised such that to
pass the stringent Solar System tests, and are compatible with the
observational data that support the concordance model.

\subsubsection{Generalizations of the Einstein-Hilbert action}

The simplest generalization consists in replacing the Ricci scalar,
$R$, in the Einstein-Hilbert action by a function $f(R)$. Note that
this gives rise to fourth-order field equations. Applying a suitable
conformal transformation of the metric, the action becomes
equivalent to a scalar-tensor theory. In detail, if we define a new
metric
$\tilde{g}_{\mu\nu}=\exp\left[\sqrt{\frac{2}{3}}\kappa\varphi\right]
g_{\mu\nu},~\kappa^2=8\pi G$, then the action becomes
\begin{equation}
S=\int\left[ \frac{1}{2\kappa^2}R[\tilde{g}]-\frac{1}{2}\tilde{g}^
{\alpha\beta}\partial_\alpha\varphi\partial_\beta\varphi -V(\varphi)
+L_{matter}\right]\sqrt{-\tilde{g}}d^4x, \label{eq:Alt10}
\end{equation}
where the potential $V$ is determined by the function $f$. With this
formulation one can, for instance, show that an arbitrary evolution
of the scale factor $a(t)$ can be obtained with an appropriate
choice of $f(R)$. It is also useful to check whether a particular
model passes Solar System tests (acceptable Brans-Dicke parameter).
One should, however, bear in mind that the two mathematically
equivalent descriptions lead to physically different properties, for
instance with regard to stability. These issues and the application
for specific functions $f$ to Friedmann spacetimes, have recently
been reviewed in \cite{NO}.

We regard such modifications as quite ad hoc. Moreover, it has not
yet been demonstrated that there are examples which satisfy all the
constraints stressed above. The same can be said on generalizations
\cite{CFT}, that include other curvature invariants, such as
$R_{\mu\nu}R^{\mu\nu},~R_{\alpha\beta\gamma\delta}R^{\alpha\beta\gamma\delta}$.
In addition, such models are in most cases \emph{unstable}, like
mechanical Lagrangian systems with higher derivatives
\cite{Wood}\footnote{This paper contains a discussion of a generic
instability of Lagrangian systems in mechanics with higher
derivatives, that was discovered by M. Ostrogradski in 1850.}. An
exception seem to be Lagrangians which are functions of $R$ and the
Gauss-Bonnet invariant $G\equiv R^2-4
R_{\mu\nu}R^{\mu\nu}+R_{\alpha\beta\gamma\delta}R^{\alpha\beta\gamma\delta}$.
By introducing two scalar fields such models can be written as an
Einstein-Hilbert term plus a particular extra piece, containing a
linear coupling to $G$. Because the Gauss-Bonnet invariant is a
total divergence the corresponding field equations are of second
order. This does, however, not guarantee that the theory is
ghost-free. In Ref.~\cite{FHT} this question was studied for a class
of models \cite{CFT} for which there exist accelerating late-time
power-law attractors and which satisfy the Solar System constraints.
It turned out that in a Friedman background there are no ghosts, but
there is instead \emph{superluminal propagation} for a wide range of
parameter space. This acausality is reminiscent of the
Velo-Zwanziger phenomenon \cite{VZ} for higher ($> 1$) spin fields
coupled to external fields. It may very well be that it can only be
avoided if very special conditions are satisfied. This issue
deserves further investigations.

\subsubsection{First-order modifications of GR}

The disadvantage of complicated fourth order equations can be
avoided by using the \emph{Palatini variational principle}, in which
the metric and the symmetric affine connection (the Christoffel
symbols $\Gamma^\alpha{}_{\mu\nu}$) are considered to be independent
fields.

It has long ago (1919) been shown by Palatini that for GR the
Palatini formulation is equivalent to the Einstein-Hilbert
variational principle, because the variational equation with respect
to $\Gamma^\alpha{}_{\mu\nu}$ implies that the affine connection has
to be the Levi-Civita connection. Things are no more that simple for
$f(R)$ models:
\begin{equation}
S=\int\left[ \frac{1}{2\kappa}f(R) +L_{matter}\right]\sqrt{-g}d^4x,
\label{eq:Alt11}
\end{equation}
where $R[g,\Gamma]=g^{\alpha\beta}
R_{\alpha\beta}[\Gamma],~R_{\alpha\beta}[\Gamma]$ being the Ricci
tensor of the independent torsionless connection $\Gamma$. The
equations of motion are in obvious notation
\begin{eqnarray}
f'(R)R_{(\mu\nu)}[\Gamma]-\frac{1}{2}f(R)g_{\mu\nu}&=&\kappa
T_{\mu\nu},\label{eq:Alt12} \\
\nabla_\alpha^{\Gamma}\left(\sqrt{-g}f'(R)g^{\mu\nu}\right)=0.
\label{eq:Alt13}
\end{eqnarray}
For the second of these equations one has to assume that
$L_{matter}$ is functionally independent of $\Gamma$. (It may,
however, contain metric covariant derivatives.)

Eq. (\ref{eq:Alt13}) implies that
\begin{equation}
\nabla_\alpha^{\Gamma}\left[\sqrt{-\hat{g}}\hat{g}^{\mu\nu}
\right]=0 \label{eq:Alt14}
\end{equation}
for the conformally equivalent metric
$\hat{g}_{\mu\nu}=f'(R)g_{\mu\nu}$. Hence, the
$\Gamma^\alpha{}_{\mu\nu}$ are equal to the Christoffel symbols for
the metric $\hat{g}_{\mu\nu}$.

The trace of (\ref{eq:Alt12}) gives
\[ Rf'(R)-2f(R)=\kappa^2 T. \]
Thanks to this algebraic equation we may regard $R$ as a function of
$T$. In the matter-free case it is identically satisfied if $f(R)$
is proportional to $R^2$. In all other cases $R$ is equal to a
constant $c$ (which is in general not unique). If $f'(c)\neq 0$, eq.
(\ref{eq:Alt13}) implies that $\Gamma$ is the Levi-Civita connection
of $g_{\mu\nu}$, and (\ref{eq:Alt12}) reduces to Einstein's vacuum
equation with a cosmological constant. In general, one can rewrite
the field equations in the form of Einstein gravity with nonstandard
matter couplings. Because of this it is, for instance,
straightforward to develop cosmological perturbation theory
\cite{Koi1}.

Koivisto \cite{Koi2} has applied this to study the resulting matter
power spectrum, and showed that the comparison with observations
leads to strong constraints. The allowed parameter space for a model
of the form $f(R)=R- \alpha R^\beta~(\alpha>0,~\beta<1)$ is reduced
to a tiny region around the $\Lambda$CDM cosmology.

The literature on this type of generalized gravity models is rapidly
increasing.

\subsubsection{Brane-world models}

Certain brane-world models\footnote{For a review, see
Ref.~\cite{Roy}.} lead to modifications of Friedmann cosmology at
very large scales. An interesting example has been proposed by
Dvali, Gabadadze and Porrati (DGP), for which the theory remains
four-dimensional at `short' distances, but crosses over to
higher-dimensional behavior of gravity at some very large distance
\cite{DGP}. This model has the same number of parameters as the
successful $\Lambda$CDM cosmology, but contains no Dark Energy. The
resulting modified Friedmann equations can give rise to universes
with accelerated expansion, due to an infrared modification of
gravity.

In Ref.~\cite{FG} the predictions of the model have been confronted
with latest supernovae data \cite{leg}, and the position of the
acoustic peak in the SDSS correlation function for a luminous red
galaxy sample \cite{Eis}. The result is that a flat DGP brane model
is ruled out at 3$\sigma$. A similar analysis was more recently
performed in \cite{Maar}, however using the SNe data \cite{rie}, but
including the CMB shift parameter that effectively determines the
first acoustic peak (see Sect.~5.1). The authors arrive at the
conclusion that the flat DGP models are within the 1$\sigma$
contours, but that the flat $\Lambda$CDM model provides a better fit
to the data. They also point out some level of uncertainty in the
use of the data, and conservatively conclude that the flat DGP
models are within joint 2$\sigma$ contours.

This nicely illustrates that observational data are restricting
theoretical speculations more and more.

The DGP models have, however, serious defects on a fundamental
level. A detailed analysis of the excitations about the
self-accelerating solution showed that there is a \emph{ghost mode}
(negative kinetic energy) \cite{GKS}. Furthermore, it has very
recently been pointed out \cite{AAH} that due to superluminal
fluctuations around non-trivial backgrounds, there is \emph{no local
causal evolution}. This infrared breakdown also happens for other
apparently consistent low-energy effective theories.

\section{Has Dark Energy been discovered in the Lab?}

It has been suggested by Beck and Mackey \cite{BM} that part of the
zero-point energy of the radiation field that is gravitationally
active can be determined from noise measurements of Josephson
junctions. This caused some widespread attention.  In a reaction we
\cite{JS} showed that there is no basis for this claim, by following
the reasoning in \cite{BM} for a much simpler model, for which it is
very obvious that the authors misinterpreted their formulae. Quite
generally, the absolute value of the zero-point energy of a quantum
mechanical system has no physical meaning when gravitational
coupling is ignored. All that is measurable are \emph{changes} of
the zero-point energy under variations of system parameters or of
external couplings, like an applied voltage. For further information
on the controversy, see \cite{BM2} and \cite{JS2}.

\begin{center}
* \quad * \quad *
\end{center}

The previous discussion should have made it clear that it is
extremely difficult to construct consistent modifications of GR that
lead to an accelerated universe at late times. The Dark Energy
problems will presumably stay with us for a long time. Understanding
the nature of DE is widely considered as one of the main goals of
cosmological research for the next decade and beyond.


\begin{thebibliography}{10}

\bibitem{leib1}
B. Leibundgut, {\it Astron. Astrophys.} {\bf 10}, 179 (2000).

\bibitem{rie}
A.~G. Riess {\it et al.}, {\it Astrophys. J.} {\bf 607}, 665 (2004);
[astro-ph/0402512].

\bibitem{p99}
S. Perlmutter {\it et al.}, {\it Astrophys. J.} {\bf 517}, 565
(1999).

\bibitem{s98}
B. Schmidt {\it et al.}, {\it Astrophys. J.} {\bf 507}, 46 (1998).

\bibitem{r98}
A.~G. Riess {\it et al.}, {\it Astron. J.} {\bf 116}, 1009 (1998).

\bibitem{leib2}
B. Leibundgut, {\it Ann. Rev. Astron. Astrophys.} {\bf 39}, 67
(2001).

\bibitem{fil04}
A.V. Filippenko, {\it Measuring and Modeling the Universe}, ed. W.L.
Freedman, Cambridge University Press, (2004); [astro-ph/0307139].

\bibitem{leg}
P. Astier {\it et al.}, astro-ph/0510447.

\bibitem{clo}
A. Clocchiatti {\it et al.}, astro-ph/0510155.

\bibitem{snap}
Snap-Homepage: http://snap.lbl.gov

\bibitem{22}
U. Seljak, and M. Zaldarriaga, Astrophys. J.\textbf{469}, 437
(1996). (See also {\it
http://www.sns.ias.edu/matiasz/CMBFAST/cmbfast.html})

\bibitem{NS8}
N. Straumann, {\it From primordial quantum fluctuations to the
anisotropies of the cosmic microwave background radiation}, to
appear in Annalen der Physik (2006); [hep-ph/0505249].

\bibitem{25}
G. Steigman, {\it Int.J.Mod.Phys.} \textbf{E15}, 1 (2006);
astro-ph/0511534.

\bibitem{26}
C.~L. Bennett {\it et al.}, ApJS \textbf{148}, 1 (2003); ApJS
\textbf{148}, 97 (2003).

\bibitem{27}
D.~N. Spergel {\it et al.}, ApJS \textbf{148} 175 (2003).

\bibitem{Sperg2}
D.~N. Spergel {\it et al.}, astro-ph/0603449.

\bibitem{Pag}
L. Page {\it et al.}, astro-ph/0603450.

\bibitem{CST}
E.~J. Copeland, M. Sami, and S. Tsujikawa, hep-th/0603057.


\bibitem{Col}
S. Cole, et al., MNRAS, \textbf{362}, 505 (2005).

\bibitem{29}
A. Blanchard, M. Douspis, M. Rowan-Robinson, and S. Sarkar, {\it
Astron.Astrophys.} \textbf{412}, 35 (2003); astro-ph/0304237.

\bibitem{BDRS}
A. Blanchard, M. Douspis, M. Rowan-Robinson, and S.
Sarkar,astro-ph/0512085.


\bibitem{Kol}
E.~W. Kolb, S. Matarrese, A. Notari, and A. Riotto, hep-th/0503117.


\bibitem{Kol2}
E.~W. Kolb, S. Matarrese, and A. Riotto, astro-ph/0506534.

\bibitem{Sas}
M. Sasaki, {\it Mon. Not. R. Astron. Soc.} \textbf{228}, 653 (1987).

\bibitem{SSS}
N. Sigiura, N. Sugiyama, and M. Sasaki, {\it Prog.Theo. Phys.}
\textbf{101}, 903 (1999).


\bibitem{Bol}
K. Bolejko, astro-ph/0512103.

\bibitem{NO}
S. Nojiri and S.~D. Odintsov, hep-th/0601213.

\bibitem{CFT}
S. Caroll, A. De Felice, V. Duvvuri, D. Easson, M. Trodden and M.
Turner, {\it Phys. Rev.} \textbf{D70}, 063513 (2005).

\bibitem{Wood}
R.~P. Woodard, astro-ph/0601672.

\bibitem{FHT}
A. De Felice, M. Hindmarsh, and M. Trodden, astro-ph/0604154.

\bibitem{VZ}
G. Velo and D. Zwanziger, {\it Phys. Rev.} \textbf{186}, 1337-41
(1969) ; {\it Phys. Rev.} \textbf{188}, 2218-22 (1969).

\bibitem{Koi1}
T. Koivisto and H. Kurki-Suonio, astro-ph/0509422.

\bibitem{Koi2}
T. Koivisto, astro-ph/0602031.

\bibitem{DGP}
G.~R. Dvali, G. Gabadadze and M. Porrati, {\it Phys. Lett.}
\textbf{B 485}, 208 (2000); hep-th/0005016.

\bibitem{Roy}
R. Maartens, {\it Living Reviews}, gr-qc/0312059.

\bibitem{FG}
M. Fairbairn and A. Goobar, astro-ph/0511029.

\bibitem{Eis}
D.~J. Eisenstein {\it et al.}, {\it Astrophys. J.} {\bf 633}, 560
(2005).

\bibitem{Maar}
R. Maartens and E. Majerotto, astro-ph/0603353.

\bibitem{GKS}
D. Gorbunov, K. Koyama, and S. Sibiryakov, {\it Phys. Rev.}
\textbf{D73}, 044016 (2006).

\bibitem{AAH}
A. Adams, N. Arkani-Hamed, S. Dubovsky, A. Nicolis, and R. Rattazzi,
hep-th/0602178.

\bibitem{BM}
C. Beck and M.~C. Mackey, {\it Phys. Lett.} \textbf{B 605}, 295
(2005); astro-ph/0406504.

\bibitem{JS}
Ph. Jetzer and N. Straumann, {\it Phys. Lett.} \textbf{B 606}, 77
(2005); [astro-ph/0411034].

\bibitem{BM2}
C. Beck and M.~C. Mackey, astro-ph/0603397.

\bibitem{JS2}
Ph. Jetzer and N. Straumann, astro-ph/0604522.
\end{thebibliography}
\end{document}